\documentclass[%
 reprint,
 amsmath,amssymb,prl,
 aps,
]{revtex4-2}

\usepackage{mathrsfs}
\usepackage{amsthm,amsmath,amssymb,lipsum}

\usepackage{graphicx}
\usepackage{dcolumn}
\usepackage{bm}

\usepackage{color}

\begin{document}

\preprint{APS/123-QED}

\title{Secondary proximity effect in a side-coupled double quantum dot structure}

\author{Jia-Ning Wang$^{1}$}
\author{Yong-Chen Xiong$^{1}$}
\email[]{xiongyc_lx@huat.edu.cn}
\author{Wang-Huai Zhou$^{1}$}
\author{Tan Peng$^{1,2}$}
\author{Ziyu Wang$^{2}$}
\email[]{zywang@whu.edu.cn}
\affiliation
{1 Hubei Key Laboratory of Energy Storage and Power Battery, and School of Mathematics, Physics and Optoelectronic Engineering, Hubei University of Automotive Technology, Shiyan 442002, People's Republic of China. \\
2 The Institute of Technological Sciences, Wuhan University, Wuhan 430072, P. R. China.}

\date{\today}

\begin{abstract}
   Semiconductor quantum dots in close proximity to superconductors may provoke localized bound states within the superconducting energy gap known as Yu-Shiba-Rusinov (YSR) state, which is a promising candidate for constructing Majorana zero modes and topological qubits. Side-coupled double quantum dot systems are ideal platforms revealing the secondary proximity effect. Numerical renormalization group calculations show that, if the central quantum dot can be treated as a noninteracting resonant level, it acts as a superconducting medium due to the ordinary proximity effect. The bound state in the side dot behaves as the case of a single impurity connected to two superconducting leads. The side dot undergoes quantum phase transitions between a singlet state and a doublet state as the Coulomb repulsion, the interdot coupling strength, or the energy level sweeps. Phase diagrams indicate that the phase boundaries could be well illustrated by $\Delta \approx c {T_{K2}}$ in all cases, with $\Delta$ is the superconducting gap, $T_{K2}$ is the side Kondo temperature and $c$ is of the order $1.0$. These findings offer valuable insights into the secondary proximity effect, and show great importance for designing superconducting quantum devices.
\end{abstract}

\maketitle


When a semiconductor quantum dot (QD) is attached by a superconductor electrode, interactions between magnetic moments in the QD and Cooper pairs in the lead \cite{Sacepe2011NaturePhysics, wang2022singlet} may result in quasiparticle excitations and low-energy bound states inside the superconducting gap, and the so-called Yu-Shiba-Rusinov (YSR) bound state or simply Shiba state could be found in the local density of states of the QD \cite{yu2005bound, shiba1968classical, rusinov1969theory}. Recently, this field has attracted significant attentions, since it provides many opportunities for systematic investigations of Majorana zero modes within which fault-tolerant quantum computation could be implemented \cite{Mourik2012Science, Prada2020NatRevPhys, Valentini2022Nature, PhysRevB.106.085420, Crawford2022npj, Cao2023SCPMA}. Furthermore, such hybrid superconductor-semiconductor systems offer unique access to design and construct superconducting quantum devices, such as topological qubits \cite{Mishra2021PRXQ, Pavesic2022PRB}, effective topological
superconductor \cite{Klinovaja2013PRL, Perge2014Science, Feldman2017NatPhys}, thermoelectric engine \cite{Germanese2022NatNano}, thermal quantum interference proximity transistor \cite{Ligato2022NatPhys}, spin-orbit coupling semiconductor nanowires \cite{xing2022reversible, kezilebieke2019observation, Mourik2012Signatures, deng2012anomalous, das2012zero}, superconducting-topological insulator hybrids \cite{Field2022Kokkeler, Chen2022superconducting, PhysRevB.106.224509}, superconducting two-dimensional (2D) devices \cite{zhou_fusion_2022, Berthold2019Observation, du2017evidence} and so on.

Basically, YSR state is induced by the proximity effect \cite{RevModPhys.77.935}, where if a superconducting material is put into contact with a nonsuperconducting one, the electron pairing correlations can propagate into the nonsuperconducting part, inducing superconducting-like properties near the interface \cite{Pillet2013PRB}. Previous works mainly concentrated on the direct proximity effect occurring on those architectures which are connected firsthand to the superconducting material \cite{trang_conversion_2020,Kezilebieke2021Electronic,schneider_proximity_2023, PhysRevB.94.140505}. However, it would be quite interesting if one nano-structure (subsystem I) is separated by a nonsuperconducting material (subsystem II) from the superconducting part. Little is known about the physical picture of the related secondary proximity effect, viz., YSR state on subsystem I.

\begin{figure}
	\vspace{-1.0em}
	\begin{center}
		\includegraphics[scale=0.25]{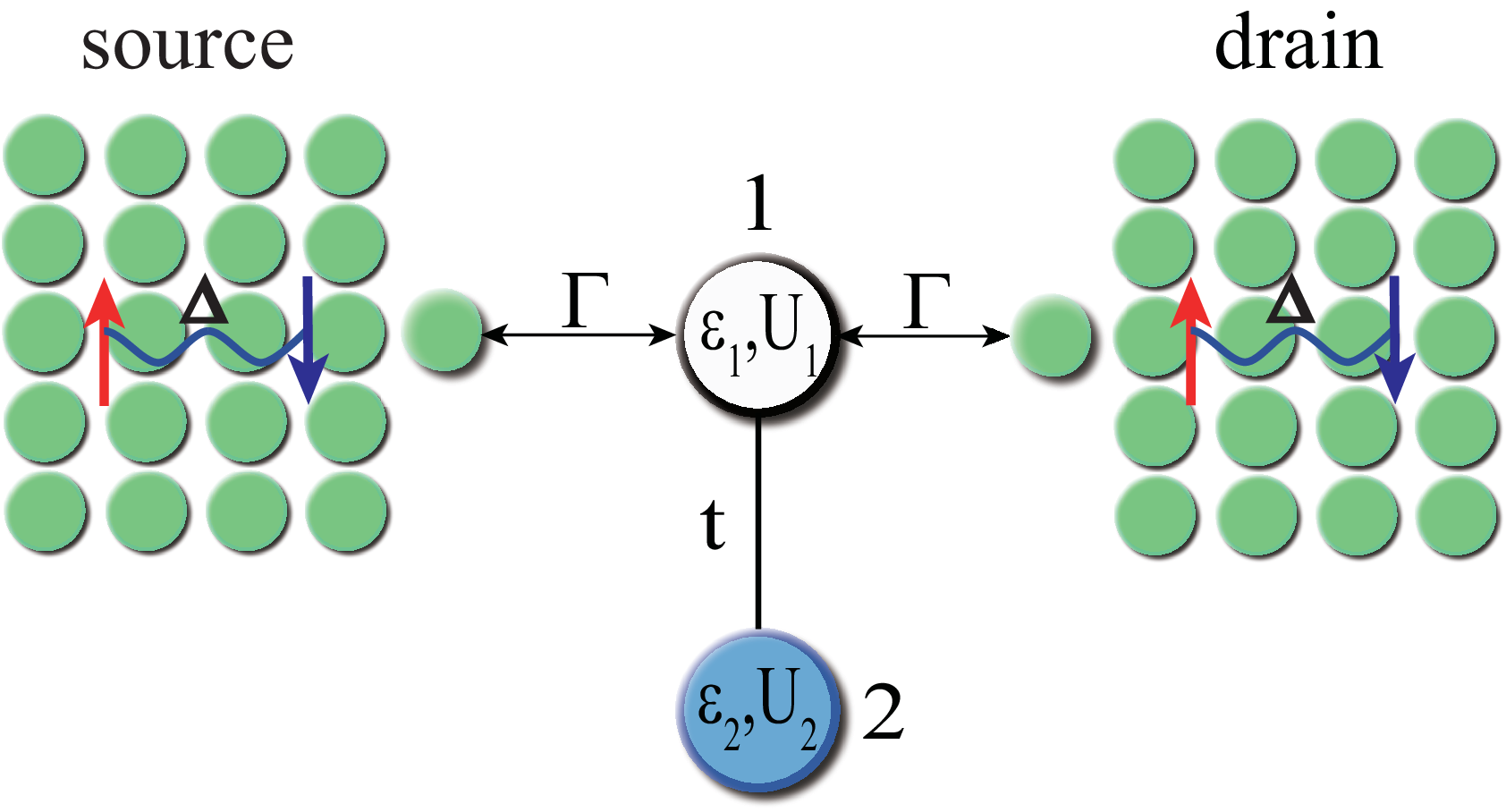}
		\caption{(Color online) Schematic illustration of the SCDQD structure connected to the superconducting electrodes. QD1 is the central QD, while QD2 is the side QD. $\varepsilon_i$ and $U_i$ are the energy levels and the on-site Coulomb repulsion of the $i$th QD, respectively. $\Gamma$ is the hybridization strength between QD1 and the superconducting electrodes. $t$ is the interdot hopping integral.  }
		\label{fig1}
	\end{center}
\end{figure}

In this work, we consider a side-coupled double quantum dot (SCDQD) device connected to two superconducting leads (see Fig.~\ref{fig1}), which is the simplest model that may present the secondary proximity effect. In this system, the central QD (QD1) is sandwiched between the superconducting source (S) and drain (D) electrodes, while the side QD (QD2) only connects directly to QD1 through interdot hopping $t$. The related second quantized Hamiltonian is given as: $\mathcal{H}$ $=$ $\sum_{\nu = S, D}{H_\nu}$ $+$ ${H_{dots}}$ $+$ ${H_{hyb}}$. Here, ${H_\nu}$ illustrates the superconducting electrodes: ${H_\nu}$ $=$ $\sum\limits_{k\sigma} \varepsilon_{\nu k\sigma } c_{\nu k\sigma }^\dag c_{\nu k\sigma}$ $-$ $\Delta \sum\limits_k (c_{k \uparrow }^\dag c_{ - k \downarrow }^\dag + H.c.)$, with $\varepsilon _{\nu k\sigma }$ is the energy level with respect to the Fermi level, $c_{\nu k\sigma }^\dag (c_{\nu k\sigma })$ is the creation (annihilate) operator for electrons with wave vector $k$, spin $\sigma$ ($=\uparrow$ or $\downarrow$), and $\Delta$ is the isotropic superconducting gap parameter. ${H_{dots}}$ is for electrons on the SCDQD:
\begin{eqnarray} \label{1}
{H_{dots}} =& \sum\limits_{i\sigma } {\varepsilon _i} d_{i\sigma }^\dag d_{i\sigma } + \sum\limits_{i\sigma } {U_i}{n_{i \uparrow }}{n_{i \downarrow }}\notag\\
          -& t\sum\limits_\sigma (d_{1\sigma }^\dag {d_{2\sigma }} + d_{1\sigma } {d_{2\sigma }^\dag}), \label{3}
\end{eqnarray}
with $\varepsilon_i$ and $U_i$ are the single electron energy and the on-site Coulomb repulsion on the $i$th ($i = 1, 2$) dot, respectively. $d_{i\sigma }^\dag $ (${d_{i\sigma }}$) creates (annihilation) a local electron on dot $i$. $n_{i\sigma} = d_{i\sigma}^{\dag}d_{i\sigma}$ is the spin$-\sigma$ number operator and $t$ is the interdot hopping integral. Finally, $H_{hyb}$ describes the coupling between QD1 and the superconducting electrodes: ${H_{hyb}} = \tau \sum\limits_{\nu k\sigma } {(c_{\nu k\sigma }^\dag {d_{1\sigma }} + H.c.)}$. Here, $\tau$ is the tunneling strength, which is assumed $k$ and $\sigma$ independent, and is symmetric with respect to the S and D electrodes.

We use the Wilson's numerical renormalization group (NRG) method \cite{PhysRevB.79.085106,PhysRevB.21.1003,li1998kondo, madhavan1998tunneling} to solve the model Hamiltonian $\mathcal{H}$. The NRG method is an unbiased nonperturbative method that works perfectly at both zero and finite temperatures, and is a quantitatively reliable technique making a close connection between theoretical and experimental studies \cite{Numerical2008Bulla, rontynen2015topological}. In our NRG calculations, we assume the density of states of a wide flat conduction band $\rho_0=1/(2\mathbb{D})$, with $\mathbb{D}$ is the half bandwidth. The hybridization function between QD1 and the electrodes could be written as $\Gamma  = \pi \rho_{0} {\tau ^2}$. We take the logarithmic discretization parameter of the leads to be $\Lambda = 3$, and keep nearly $3000$ low-lying states within each iteration step. At temperature $T$, the local density of states (LDOS) of each QD can be defined as:
\begin{eqnarray}\label{2}
{A_i}(\omega ,T) =  - \frac{1}{\pi }{\mathop{\rm Im}\nolimits} {G_i}(\omega  + i\delta ).
\end{eqnarray}
Here, ${G_i}(\omega  + i\delta ) = {\left\langle {\left\langle {{d_{i\sigma }};d_{i\sigma }^\dag } \right\rangle } \right\rangle _{\omega  + i\delta }}$ is the Green's function. In the following, we abbreviate $A_{i}(\omega, T)$ at zero temperature as $A_{i}(\omega)$.

\begin{figure}
 	\begin{center}
 		\includegraphics[scale=0.28]{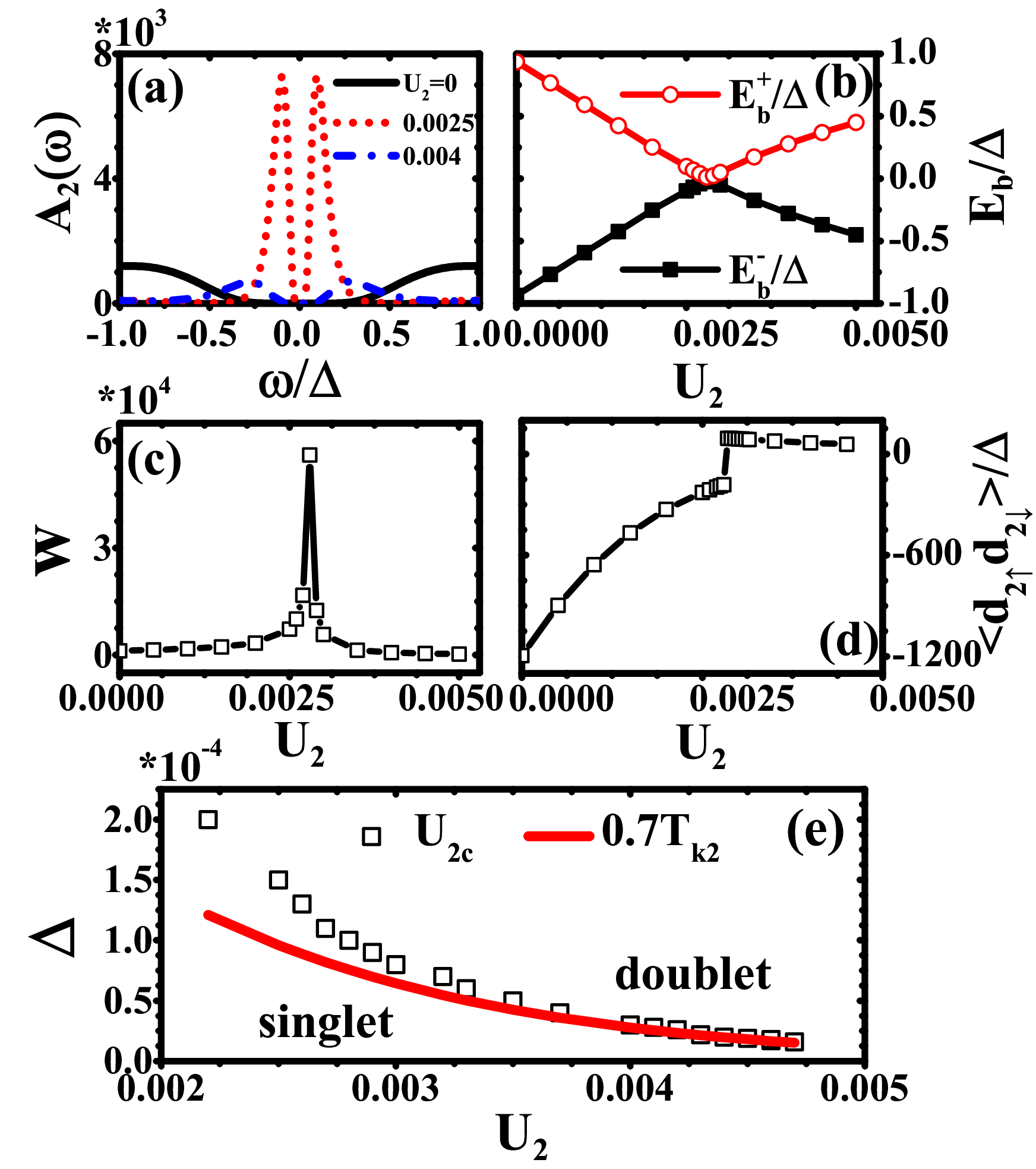}
 		\caption{(Color online)
 			(a) The LDOS of electrons in QD2 at nearly zero temperature $A_2(\omega)$ in the gap region [$-\Delta$, $\Delta$] for various $U_2$.
 			(b) The energies $E_b^{\pm}$ and (c) weights $W$ of YSR peaks as functions of $U_2$. $E_b^{\pm}$ are scaled by $\Delta$. (d) The expectation value of superconducting order in QD2 ${\left\langle {{d_{2 \uparrow }}{d_{2 \downarrow }}} \right\rangle / \Delta }$. (e) Phase diagram of the singlet and doublet states as a function of $U_2$ and $\Delta$. The empty square describes the critical points $U_{2c}$ for fixed $\Delta$. The red line indicates the fitting function Eq.~(\ref{5}). The other parameters are given by $U_1 = 0$, $t = 0.002$, $\Gamma = 0.01$, $\Delta = 0.0001$, and $\varepsilon_i = -U_i/2$.}
 		\label{fig2}
 	\end{center}
\end{figure}

In Fig.~\ref{fig2}(a), we show the LDOS of QD2 $A_2(\omega)$ in the gap regime [$-\Delta$, $\Delta$] for fixed $U_1 = 0$, $t = 0.002$, $\Delta = 0.0001$ and various $U_{2}$. Here, we have chosen $\mathbb{D}$ as the energy unit. For $U_{i} = 0$, there exists an obvious gap exists in $A_1(\omega)$, with its edges located at $\pm \Delta$. Meanwhile, a pair of symmetric YSR states could be found in $A_2(\omega)$ for each $U_{2}$. As $U_2$ increases, the YSR peaks, viz. the energies of YSR bound state $E_b^{+}$ and $E_b^{-}$, move closer to the Fermi surface firstly if $U_{2} < U_{2c}$, then toward the gap edge when $U_{2} > U_{2c}$. Correspondingly, the weights of the YSR peaks $W$ increase till $U_{2c}$, and then they decrease gradually. The whole pictures of $E_b^{\pm}$ and $W$ varying with $U_2$ are plotted in Figs.~\ref{fig2}(b) and \ref{fig2}(c) respectively.

The above phenomenon indicates a quantum phase transition (QPT) at the critical point ${U_{2c}} \approx 0.0028$, which finds its counterpart in the expectation value of superconducting order of QD2 ${\left\langle {{d_{2 \uparrow }}{d_{2 \downarrow }}} \right\rangle }$ shown in Fig.~\ref{fig2}(d). It is seen that $\left| {\left\langle {{d_{2 \uparrow }}{d_{2 \downarrow }}} \right\rangle } \right|$ is large for $U_{2} < U_{2c}$, since the YSR state can be considered as a linear combination of the empty and doubly occupied states \cite{Bauer2007JPCM}. When $U_{2}$ exceeds $U_{2c}$, $\left| {\left\langle {{d_{2 \uparrow }}{d_{2 \downarrow }}} \right\rangle } \right|$ drops to a smaller value, for $U_{2}$ favors the singly occupied state in QD2, hence the superconductivity of QD2 is suppressed.

One can naturally interpret the evolution of the above YSR bound state by considering the energy difference between the ground state and the low excited states of the whole SCDQD system. However, we highlight that since QD1 acts as another new superconducting medium with an energy gap $\Delta$ connected directly to QD2 due to the typical proximity effect from the original S and D electrodes, the above QPT could then be attributed to the competition between superconducting state and side Kondo behavior \cite{Silva2006PRL, Wang2022Unified} on QD2. The relevant energy scales are the isotropic superconducting gap $\Delta$ and the side Kondo temperature of QD2 $T_{K2}$. Here $T_{K2}$ can be captured by \cite{Haldane1978Theory, Wang2022Unified}:
  \begin{eqnarray}\label{3}
    {T_{K2}} = {U_2}\sqrt {\rho {J_2}} {e^{ - 1/(\rho {J_2})}},
  \end{eqnarray}
with $\rho {J_2} = 8{\Gamma _{c - s}}/\pi {U_2}$ is the effective Kondo coupling between QD1 and QD2. For parameters given in our present model, the effective hybridization function $\Gamma _{c - s}$ between them can be expressed as \cite{Wang2022Unified}: ${\Gamma _{c - s}} = \pi A_1^0(\omega ){t^2}$. Here, $A_1^0(\omega )$ is the LDOS of QD1 with $t = 0$ and normal conduction leads. With the aid of the Green's function, $A_1^0(\omega )$ could be written as:
  \begin{eqnarray}\label{4}
    A_1^0(\omega ) = \frac{\Gamma }{{\pi [{{(\omega  - {\varepsilon _1})}^2} + {\Gamma ^2}]}}.
  \end{eqnarray}
When ${U_2}$ is small, such that ${T_{K2}} > \Delta $, we have a singlet ground state of QD2 ($S_{2} = 0$). Whereas if $U_{2}$ is large enough with ${T_{K2}} < \Delta $, the ground state is a spin doublet ($S_{2} = 1/2$).

Summarizing the behaviour for different superconducting gaps $\Delta$, we then obtain a phase diagram for the singlet and doublet states in Fig.~\ref{fig2}(e). It is found that the ground state of QD2 is always a singlet state when $U_2$ is small. As $U_2$ increases, it becomes a doublet state. The critical value $U_{2c}$ increases gradually with decreasing $\Delta$. Quite interestingly, in the strong interaction region, the phase boundary relationship between the singlet and doublet ground states could be given by:
  \begin{eqnarray}\label{5}
    \Delta = 0.7 {T_{K2}}.
  \end{eqnarray}
viz., for $T_{K2}/\Delta > 0.7$, we have a singlet ground state, while for $T_{K2}/\Delta < 0.7$ the ground state is a doublet. However, for smaller $U_{2}$, the estimated line deviates from the numerical results. These results are similar to those for the single impurity case \cite{Yoshioka2000JPSJ, Bauer2007JPCM}.

\begin{figure}
	\begin{center}
		\includegraphics[scale=0.28]{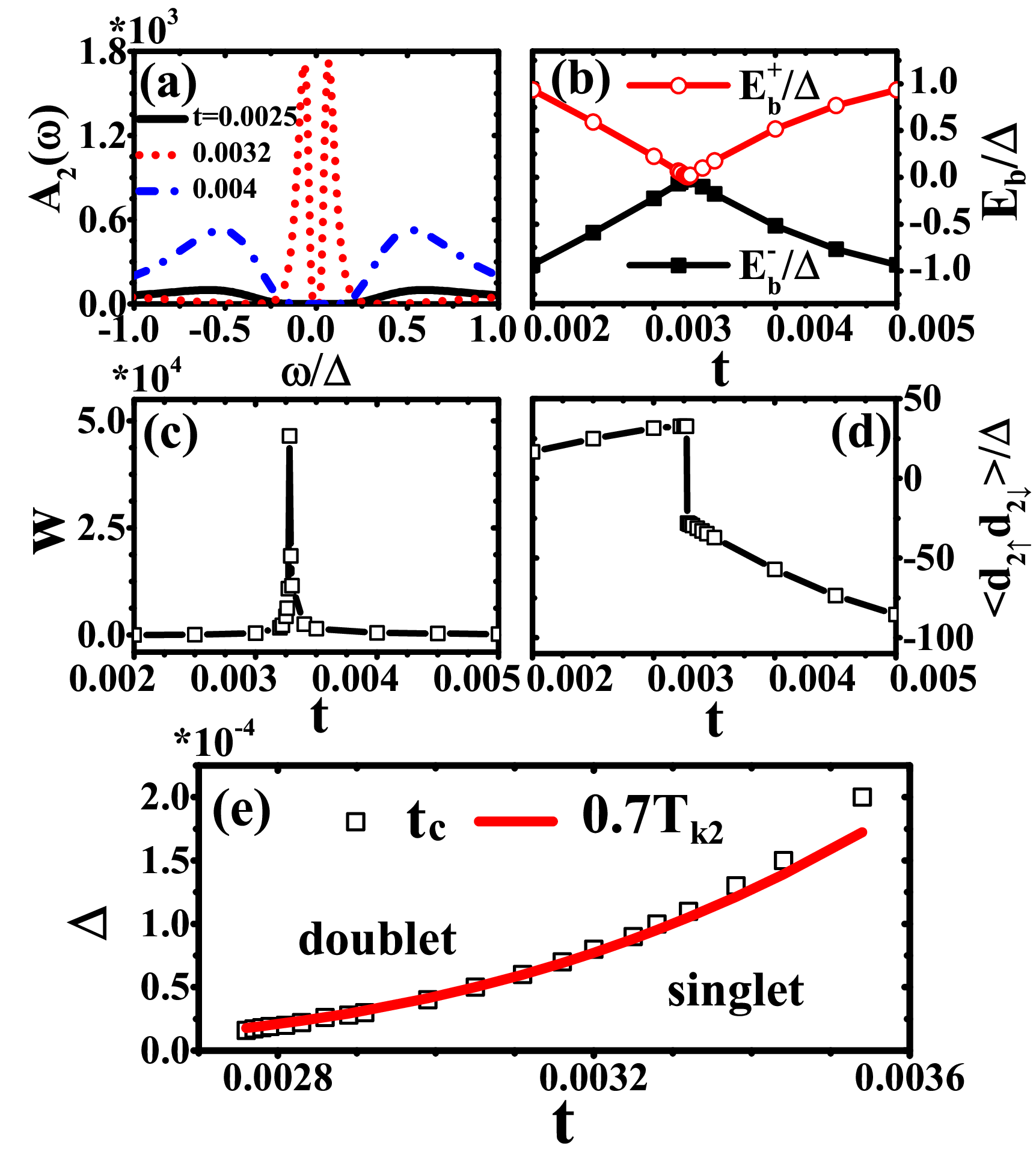}
		\caption{(Color online) (a) $A_2(\omega)$ in the gap region [$-\Delta$, $\Delta$] for various $t$ with $U_{2} = 0.01$ and $\Delta = 0.0001$. (b) $E_b^{\pm}/\Delta$, (c) $W$, and (d) ${\left\langle {{d_{2 \uparrow }}{d_{2 \downarrow }}} \right\rangle / \Delta }$ as functions of $t$. (e) Phase diagram of the singlet and doublet states in the $\Delta - t$ plane. The empty squares describe the critical points $t_{c}$ for fixed $\Delta$. The red line indicates the fitting function Eq.~(\ref{5}). The other parameters are given the same as Fig.~\ref{fig2} unless specifically stated.}
		\label{fig3}
	\end{center}
\end{figure}

In the aforementioned case, we mainly analyze the secondary proximity effect in QD2 affected by $U_2$. Now we turn to the case by varying $t$, with fixed $U_1 = 0$, $U_2 = 0.01$ and $\Delta = 0.0001$ in most instances. From the LDOS of QD2 $A_{2}(\omega)$ in Fig.~\ref{fig3}(a), it can be seen that the YSR peak firstly moves away from the gap edge to the Fermi surface as $t$ sweeps upwards, and reapproaches the gap edge afterwards. This behavior is well illustrated by Fig.~\ref{fig3}(b). For small $t$, the YSR peaks are located at the edge of the gap. Then they move gradually towards the center if $t$ turns up. Close to the critical value $t_c$, the ground state switches from a doublet state $(S_2 = 1/2)$ to a singlet state $(S_2 = 0)$, $E_b^ + $ and $E_b^ - $ cross at $t = t_c$. When $t$ is further increased, the YSR peaks move towards the gap edge again and gradually hold at there. In Fig.~\ref{fig3}(c), we can observe that the weight of YSR peak show a tendency to increase and then decrease with the increasing $t$, which reaches a maximum at $t_c \approx 0.00328$.

Fig.~\ref{fig3}(d) depicts ${\left\langle {{d_{2 \uparrow }}{d_{2 \downarrow }}} \right\rangle }/\Delta$ as a function of $t$. It is seen that $\left| {\left\langle {{d_{2 \uparrow }}{d_{2 \downarrow }}} \right\rangle } \right|$ gradually strengthens as $t$ increases. Because if $t$ is applied, the particle-hole (p-h) symmetry of both dots is broken, and the probability of the empty or fully occupied states on QD2 increases. When $t$ exceeds $t_{c}$, $ {\left\langle {{d_{2 \uparrow }}{d_{2 \downarrow }}} \right\rangle } $ changes abruptly to a negative value, with $\left| {\left\langle {{d_{2 \uparrow }}{d_{2 \downarrow }}} \right\rangle } \right|/\Delta$ is enhanced. In this process, $J_{2}$ grows progressively with increasing $t$, resulting in an enlargement of the Kondo temperature $T_{K2}$ as per Eq.~(\ref{3}). If $T_{K2}$ overwhelms $\Delta$, the binding energy of the Kondo singlet between QD2 and QD1 is favored.
So the side Kondo behavior is dominant with the ground state turns to be the singlet $S_{2} = 0$ from the doublet $S_{2} = 1/2$.

The related phase diagram is plotted in Fig.~\ref{fig3}(e). For small $t$, the ground state of QD2 is always a doublet, whereas if $t$ is large enough, the ground state turns to be a singlet through a QPT. The critical point $t_{c}$ increases if $\Delta$ is lifted up. One observes that $t_{c}$ could also be well illustrated by Eq.~(\ref{5}). Viz., when $\Delta/{T_{K2}} > 0.7$, the singlet bound state known as Cooper pair in the QD1 is favored, and the QD2 is in a localized doublet state. Whereas if $\Delta/{T_{K2}} < 0.7$, the Kondo singlet between two dots is dominant, the side Kondo behavior overwhelms, and the ground state of QD2 changes to a spin singlet.

\begin{figure}
	\begin{center}
		\includegraphics[scale=0.40]{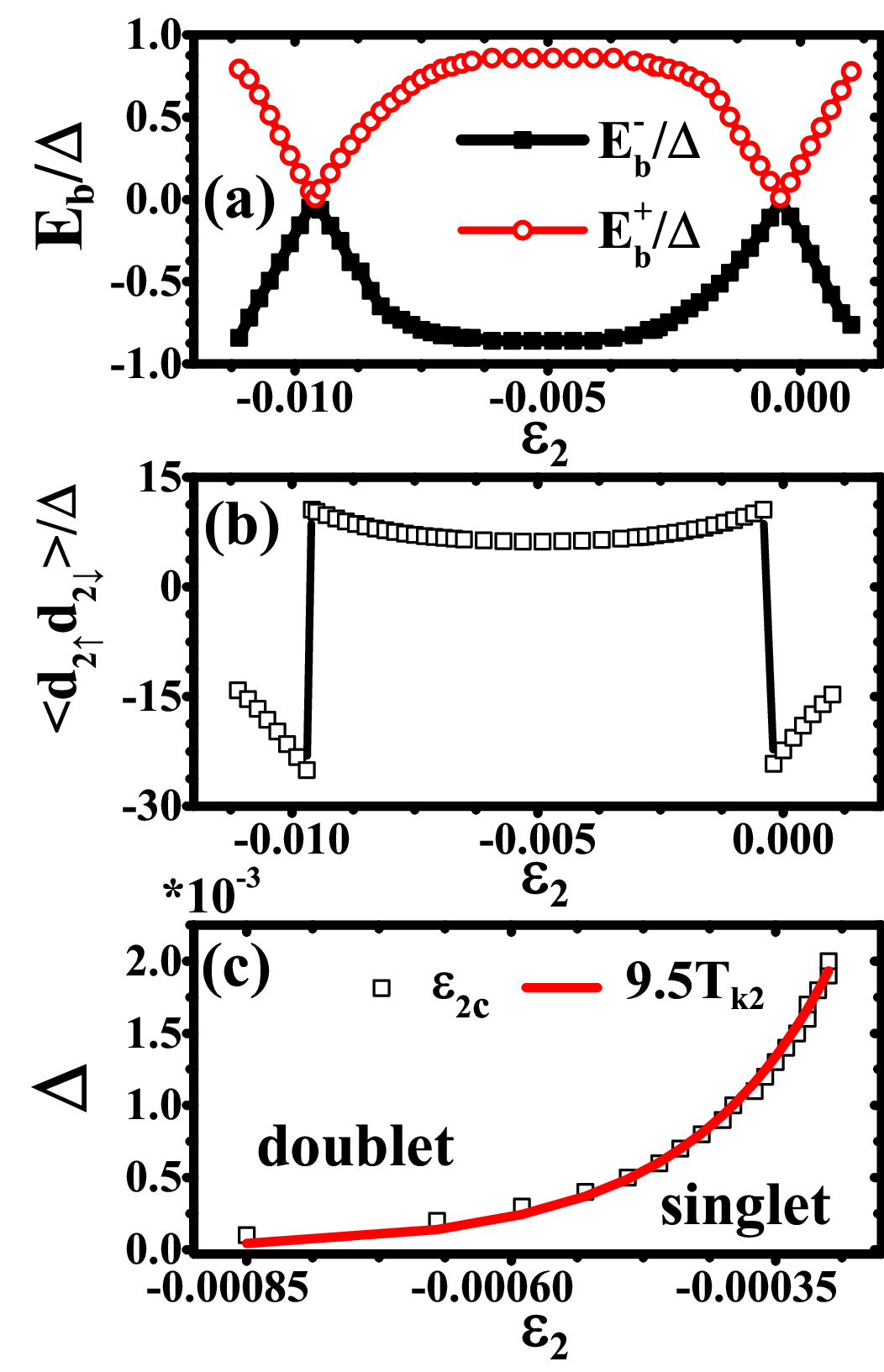}
		\caption{(Color online)
		(a) $E_b^{\pm}/\Delta$ and (b) ${\left\langle {{d_{2 \uparrow }}{d_{2 \downarrow }}} \right\rangle / \Delta }$ as functions of $\varepsilon_{2}$ with $U_2=0.01$, $t=0.002$ and $\Delta=0.001$. (c) Phase diagram of the singlet and doublet states in the $\Delta - \varepsilon_{2}$ plane. The empty squares describe the critical points in the right side $\varepsilon_{c2}$ for fixed $\Delta$. The red line indicates the fitting function Eq.~(\ref{6}). The other parameters are given the same as Fig.~\ref{fig2} unless specifically stated.}
		\label{fig4}
	\end{center}
\end{figure}

In Fig.~\ref{fig4}(a), we depict the energy dependence of YSR bound states on ${\varepsilon _2}$ with $U_2=0.01,$ $t=0.002$ and $\Delta=0.001$. It can be observed that a pair of YSR resonances appear within the superconducting gap. Due to the p-h symmetry, we only focus on the case ${\varepsilon _2} \geq -U_2/2$ in the following discussion. As ${\varepsilon _2}$ increases, the energy of the bound states $\left| E_b^{\pm} \right|$ first decreases toward zero, then gradually increases, indicating a QPT at about ${\varepsilon _2} = 0$. Such a QPT could also be found in $\left| {\left\langle {{d_{2 \uparrow }}{d_{2 \downarrow }}} \right\rangle } \right|$, cf., Fig.~\ref{fig4}(b). The underlying physical picture for the above phenomenon lies in the following. For parameters given in such a case, and ${\varepsilon _2}$ is in the singly occupied regime (${\varepsilon _2} \sim -U_{2}/2$), the ground state of QD2 is a doublet state with $S_{2} = 1/2$. When ${\varepsilon _2}$ is increased, the ground state turns to be a singlet due to QD2 is almost empty.

The phase diagram affected by ${\varepsilon_2}$ is shown in Fig.~\ref{fig4}(c), with a symmetric one occurring around ${\varepsilon _2} = -U_{2}$, but is not given here. When $\varepsilon_2$ is small, satisfying $-U_{2}/2 \leq \varepsilon_{2} < \varepsilon_{2c}$, the ground state is always a doublet state. However, as $\varepsilon_2$ becomes sufficiently large, the ground state undergoes a QPT and transfers into a singlet state, with the critical point $\varepsilon_{2C}$  increases for larger $\Delta$. The boundary between the singlet and doublet ground states can be described in terms of $\Delta$ and $T_{K2}$, which can be fitted by
\begin{eqnarray}\label{6}
  \Delta = 9.5 {T_{K2}}.
\end{eqnarray}
Here, $T_{K2}$ is described by \cite{Haldane1978Theory, Scott2010Nano}:
\begin{eqnarray}\label{7}
    {T_{K2}} = \sqrt {{U_2}{\Gamma _{c - s}}} \exp \left[\frac{{\pi {\varepsilon _2}({\varepsilon _2} + {U_2})}}{{{U_2}{\Gamma _{c - s}}}}\right]
\end{eqnarray}
One finds Eq.~(\ref{7}) is consistent with our numerical results.

In summary, we have provided an in-depth investigation of the YSR states and the QPTs in a side-coupled double quantum dot devices, with QD1 connected directly to two superconducting leads. We have shown that the SCDQD system can be tailored to explore the secondary proximity effect. If QD1 can be treated as a noninteracting resonant level, it triggers an energy gap whose width is nearly the same as the superconducting leads. The YSR peaks could be found in QD2 due to the secondary proximity effect. In such a case, QD1 could be considered as another superconducting lead, and the bound state in QD2 behaves similarly to the case of a single impurity connected to two superconducting leads. Specifically, the ground state of QD2 undergoes a transformation from a singlet state to a doublet state as the Coulomb repulsion $U_2$ sweeps. On the other hand, as the interdot coupling strength $t$ is adjusted, QD2 transforms vise versa. In addition, when changing the energy level $\varepsilon_2$, QD2 experiences QPTs from a singlet state to a doublet state and then to a singlet state again. Phase diagrams in the $\Delta - U_{2}$, $\Delta - t$ and $\Delta - \varepsilon_2$ planes have also been demonstrated, showing that the phase boundaries could be well fitted by $\Delta \approx c {T_{K2}}$ in all the above cases, with $c$ is a fitting parameter of order $1.0$, similar to the single impurity case with $c = 0.3$. These findings may show great importance for the design and application of superconducting devices and provide new ideas for further related works.

This work is financially supported by the Natural Science Foundation of Hubei Province under grant Nos.~2023AFB456 and 2023AFB891, the Foundation of Hubei Educational Committee under grant Nos.~Q20161803 and Q20221803, and the Science and Technology Innovation Team Research Fund (Institute of Shiyan Industrial Technology of Chinese Academy of Engineering) under grant No.~ZCTD202201.

\end{document}